\documentclass[5p,twocolumn,times,number]{elsarticle}
\usepackage{amsmath}
\usepackage{amssymb}
\usepackage{graphics}
\usepackage{graphicx}
\usepackage{epsfig}
\usepackage{dcolumn}
\usepackage{bm}
\usepackage{lineno}
\begin{document}
	\begin{frontmatter}
		
		
		\title{Effect of relative humidity on the long-term operation of a \\single mask triple GEM chamber}
		
		\author[]{S.~Chatterjee\corref{cor}}
		
		\ead{sayakchatterjee@jcbose.ac.in, sayakchatterjee896@gmail.com}
		
		\author[]{A.~Sen}
		\author[]{S.~Das}
		\author[]{S.~Biswas}
		
		\cortext[cor]{Corresponding author}
		\address[]{Department of Physics, Bose Institute, EN-80, Sector V, Kolkata-700091, India}

		\begin{abstract}
			
			The characteristic studies of a Single Mask~(SM) triple Gas Electron Multiplier (GEM) detector are carried out using Ar/CO$_2$ gas mixture in 70/30 volume ratio in continuous flow mode. A Fe$^{55}$ X-ray source is used for this work. The gain and energy resolution are studied from the 5.9 keV X-ray energy spectra. 
			After correcting the effects of temperature and pressure variation, the effect of relative humidity on the gain and energy resolution of the chamber is investigated. The details of the experimental setup, measurement methods and results are presented in this article.
		\end{abstract}
		
		\begin{keyword}
			Single Mask GEM \sep Gas detector \sep Gain \sep Energy resolution \sep Relative humidity
			
		\end{keyword}
	\end{frontmatter}
	
	\section{Introduction}\label{intro}
	
	Good position resolution~($\sim$~70~$\mu$m) and high rate handling capability~($\sim$~1~MHz/mm$^2$) make the GEM detector one of the advanced members of the Micro Pattern Gas Detector~(MPGD) group and is being used in many High Energy Physics~(HEP) experiments as a tracking device~\cite{ref1}. A proper understanding of the performance under long-term irradiation is an essential criterion for any detector used in HEP experiment~\cite{ref4,ref5}. In this study, the effect of temperature, pressure and relative humidity~(RH) variation on the performance of a Single Mask~(SM) triple GEM chamber is investigated under continuous irradiation with 5.9 keV X-ray from a Fe$^{55}$ radioactive source of activity $\sim$~20~mCi. 
	The chamber is operated with Ar/CO$_2$ gas mixture in a 70/30 volume ratio in a continuous flow mode. Conventional NIM electronics are used for detector biasing and data acquisition. 
	
	
	\section{Detector descriptions and experimental set-up}\label{setup}
	
	A SM triple GEM chamber of dimension 10~cm~$\times$~10~cm is used for this study. The drift gap, transfer gaps and induction gap are kept at 3~mm, 2~mm and 2~mm respectively. The biasing of the GEM foils is done by using a resistive chain as discussed in Ref.~\cite{ref9}. A filter circuit is introduced between the high voltage~(HV) line and the resistive chain to bypass the ac components. The readout board of the chamber consists of 256 XY tracks. For this study, the sum-up signal is used instead of using the individual pad readout. Four sum-up boards are used in this chamber. 
	The signal from the GEM chamber is fed to a charge sensitive preamplifier having a gain of 2~mV/fC and a shaping time of 300~ns~\cite{Preamp}. The output of the preamplifier is fed to a linear Fan-In Fan-Out~(FIFO) module to create an identical analog signal at the output. One output from FIFO is fed to a Single Channel Analyzer~(SCA), the output of which above the noise threshold are counted using the NIM scaler. 
	Another output of the FIFO is fed to a Multi-Channel Analyzer~(MCA) for storing the X-ray spectra on the PC. The gain and energy resolution of the chamber is calculated by fitting the MCA spectra using a Gaussian distribution. The schematic of the electronic circuit diagram and method of calculating the gain and energy resolution of the chamber is discussed in Ref.~\cite{ref5}. The ambient temperature, pressure and relative humidity are stored using a data logger built-in house~\cite{ref6}.  
	
	\section{Results}\label{res}
	
\begin{figure}[htb!]
		\begin{center}
			\includegraphics[scale=0.25]{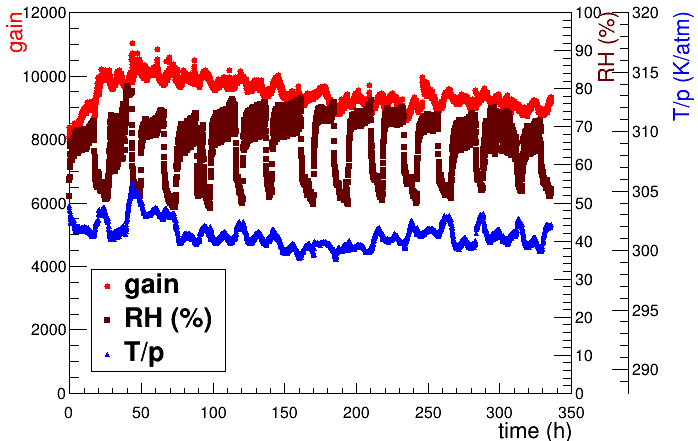}
			\includegraphics[scale=0.25]{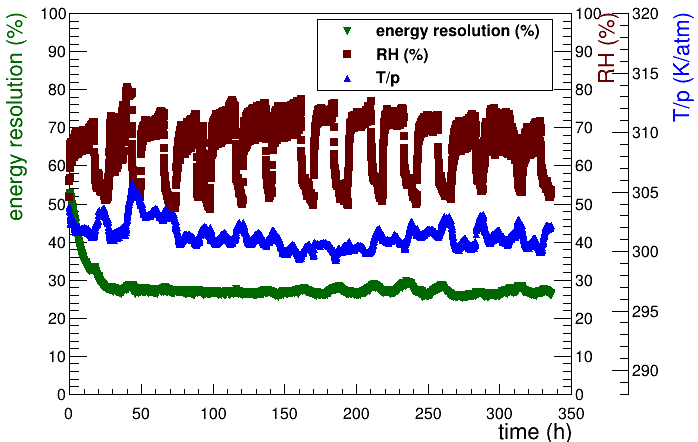}
			\vspace{-0.25cm} 
			\caption{Variation of gain (top), energy resolution (bottom), T/p and RH as a function of time (colour online).}
			\vspace{-1.1cm} 
			\label{fig3}
		\end{center}
	\end{figure}

\begin{figure}[htb!]
		\begin{center}
			\vspace*{0.4cm}			
			\includegraphics[scale=0.25]{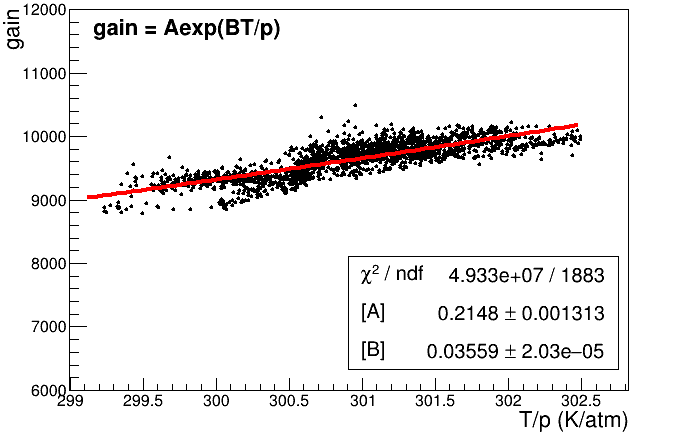}
			\includegraphics[scale=0.25]{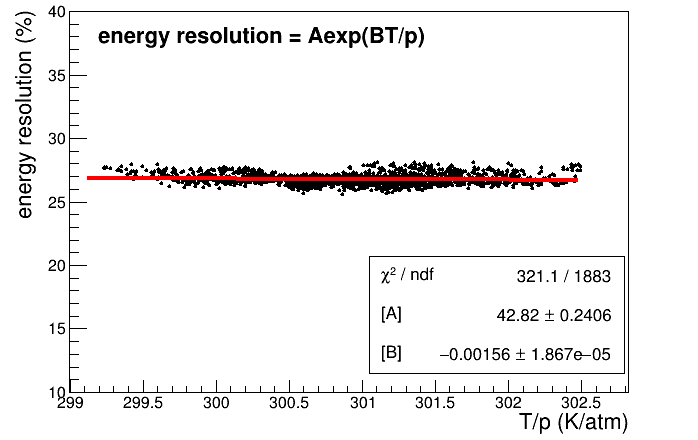}
			\vspace{-0.35cm} 
			\caption{Correlation of gain~(top) and energy resolution~(bottom) with T/p (colour online).}
			\vspace{-0.80cm} 			
			\label{fig4}
		\end{center}
	\end{figure}

\begin{figure}[htb!]
		\begin{center}
			\includegraphics[scale=0.25]{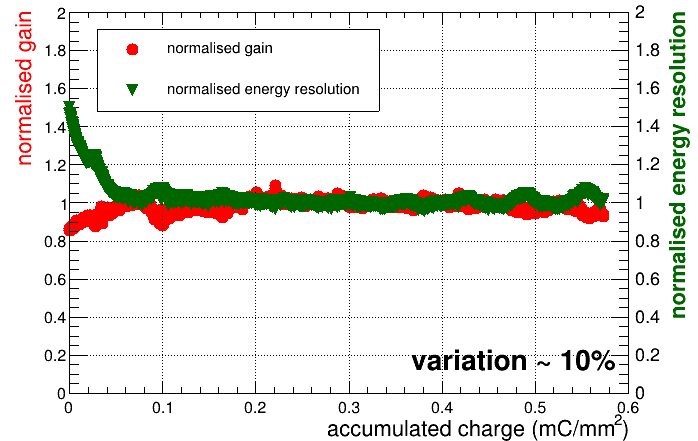}
			\caption{Variation of normalised gain and energy resolution as a function of accumulated charge (colour online). 
			}
			\vspace{-0.80cm}
			\label{fig5}
		\end{center}
	\end{figure}

	The voltage across each of the GEM foils is kept at $\sim$~405~V and the chamber is irradiated continuously with Fe$^{55}$ X-ray at a rate of $\sim$~2~kHz/mm$^2$. It is well known that the gain of any gaseous detector increases with increasing temperature~(T~=~t~+~273~K) and decreases with increasing pressure (p~atm), or in other words, the gain of the chamber is found to be positively correlated with T/p variation~\cite{ref7}. It is expected that the energy resolution will improve with the increased gain of the chamber, therefore the energy resolution is expected to be negatively correlated with T/p variations. In Fig.~\ref{fig3}, the variation of gain, energy resolution, T/p and RH is shown as a function of time.
	The correlation of gain and energy resolution with the temperature to pressure (T/p) ratio is shown in Fig.~\ref{fig4}. The gain and energy resolution is parameterized by an exponential function of the form $\it Aexp(BT/p)$, where $\it A$ and $\it B$ are the free parameters. 
	The gain and energy resolution of the chamber is normalised using the parameters obtained from the correlation curves to eliminate the effects of T/p variations. The variation of normalised gain and energy resolution is shown in Fig.~\ref{fig5} as a function of the accumulated charge per unit area. The method of calculating the accumulated charge per unit area is taken from Ref.~\cite{ref5}.
	The initial decrease in the normalised gain and increase in the normalised energy resolution is due to the charging-up effect of the chamber~\cite{ref9,ref8}. 
	
\begin{figure}[htb!]
		\begin{center}
			\includegraphics[scale=0.25]{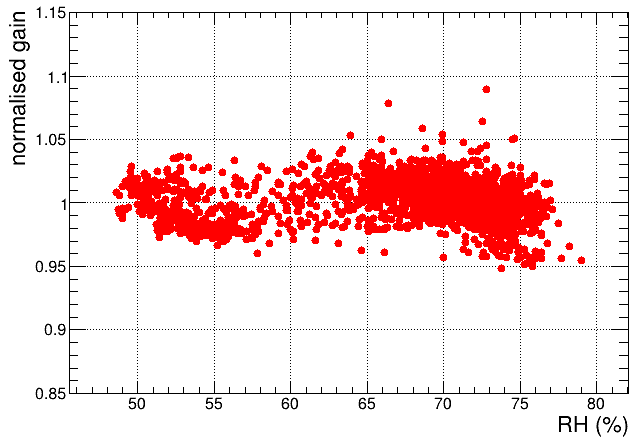}
			\includegraphics[scale=0.25]{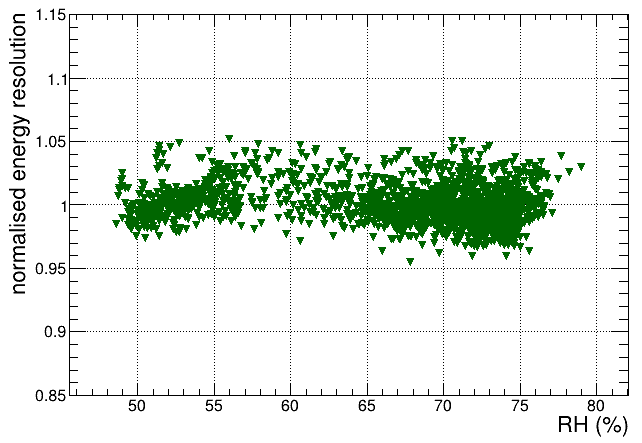}
			\vspace{-0.2cm} 
			\caption{Variation of normalised gain~(top) and energy resolution~(bottom) as a function of RH (colour online).}
			\vspace{-0.6cm}
			\label{fig6}
		\end{center}
	\end{figure}

	To study the effect of RH on the performance of the chamber, the T/p normalised gain and energy resolution of the chamber is plotted as a function of RH and shown in Fig.~\ref{fig6}. 
	No significant correlation is observed between the T/p normalised gain and energy resolution with RH~\cite{ref10}.  
	
	\vspace*{-.250cm}
	\section{Summary}
	
	The characterisation of a SM triple GEM chamber is performed in terms of its gain and energy resolution variation as a function of temperature, pressure and humidity
	with Ar/CO$_2$ gas mixture in a 70/30 volume ratio. The effect of relative humidity on the performance of the chamber is investigated after eliminating the effects of T/p variation on the gain and energy resolution of the chamber. No significant correlation is observed between T/p normalised gain and energy resolution with the relative humidity.
	
	\vspace*{-.450cm}
	\section{Acknowledgements}
	\vspace{-0.10cm}
	
	The authors would like to thank Prof. S. K. Ghosh for valuable discussions in the course of this study. The authors would also like to thank Ms. Shreya Roy for writing the initial analysis code. This work is partially supported by the research grant SR/MF/PS-01/2014-BI from DST, Govt. of India, and the research grant of the CBM-MuCh project from BI-IFCC, DST, Govt. of India. 

\end{document}